\documentclass[epj,nopacs]{svjour}
\usepackage{epsfig}
\usepackage{graphicx}
\usepackage{graphics}
\usepackage{subfigure}
\usepackage{multirow}
\ifx\pdfoutput\undefined
\usepackage[dvips]{hyperref}
\else
\usepackage[pdftex]{hyperref}
\pdfpagewidth=\paperwidth
\pdfpageheight=\paperheight
\fi 
\hypersetup{colorlinks=false,breaklinks=true,plainpages=false}

\begin{document}
\title{Electroweak boson detection in the ALICE muon spectrometer}
\subtitle{W and Z studies in hadron-hadron collisions at LHC}

\author{Z. Conesa del Valle\inst{} \thanks{Joint thesis supervision between the Universitat Aut\`onoma de Barcelona and the Universit\'e de Nantes.} 
  for the ALICE Collaboration
}
\authorrunning{Z. Conesa}
%
%
\institute{SUBATECH (\'Ecole des Mines de Nantes, CNRS/IN2P3, Universit\'e de Nantes), Nantes, France}
\date{Received: August 8, 2006}
%
\abstract{
The ALICE capabilities for W and Z detection at LHC are discussed. The contributions to single muon transverse momentum distribution are evaluated. The expected performance of the muon spectrometer for detecting W and Z bosons via their muonic decay is presented. 
Possible application for the study of in-medium effects in ultrarelativistic heavy ion collisions is discussed.
\PACS{
      {PACS-key}{discribing text of that key}   \and
      {PACS-key}{discribing text of that key}
     } 
} 
\maketitle
\section{Introduction}
\label{intro}

Electroweak bosons properties have been studied at LEP (CERN), SLC (SLAC) and Tevatron (FNAL) colliders in $p \bar{p}$ and $e^+ e^-$ collisions \cite{PDGBook}. 
At the LHC, large energy will be available in the center-of-mass, enabling the possibility to produce W and Z bosons in p-p collisions, and in nucleus-nucleus (A-A) collisions (see production cross-sections in table \ref{tab:cross_section}).
\newline
W and Z bosons are massive particles produced in initial hard collisions. 
In the lowest order approximation, they are produced by the quark ($q$) - antiquark ($\bar{q}$) annihilation process:
\begin{displaymath}
q~+~\bar{q'}~\rightarrow~W^{\pm}~; \qquad   q~ +~\bar{q}~\rightarrow~Z~.
\end{displaymath}
These subprocesses are characterized by the scale $Q^2 = M^2$ and the Bjorken-$x$ values, which can be determined by $x_{1,2} \sim \frac{M}{\sqrt{s}} \, e^{\pm y}$ \cite{tricoli,copper}, where $M$ is the mass of the electroweak boson, $\sqrt{s}$ is the center-of-mass energy of the nucleon-nucleon collision, and $y$ is the rapidity.
\newline
Electroweak boson measurements at LHC will provide important information: 
\begin{itemize}
\item[-] These subprocesses are considered as Standard Model benchmarks. Their production cross-sections are 'known' with a precision dependent on the parton distribution functions (PDFs) uncertainties. Therefore, they have been suggested as 'standard-candles' for luminosity measurements, and to improve the evaluation of the detector performances \cite{tricoli,copper}. 
\item[-] In proton-proton (p-p) collisions, they will be sensitive to the quark PDFs at high $Q^2$ ($Q = M_{W/Z}$). 
\item[-] In proton-nucleus (p-A) collisions, quark nuclear modification effects will become accessible at the same scale.
\item[-] Since electroweak bosons are probes produced in hard primary collisions and they do not interact strongly with the surrounding medium created in the collision, they will allow binary scaling cross-checks in A-A collisions. They could then be used as a reference for observing medium induced effects on other probes, like the suppression of high transverse momentum ($p_T$) muons from charm and beauty. 
\end{itemize}
W/Z bosons can be detected via their leptonic decay. The decay $W^+~\rightarrow~l^{+}~\nu_{l}$ ($W^-~\rightarrow~l^{-}~\bar{\nu_{l}}$) has a branching ratio of $10.7~\%$, and the decay $Z~\rightarrow~l^{+}~l^{-}$ has a branching ratio of $3.37~\%$ \cite{PDGBook}.
\newline
In this paper, W/Z production at LHC is discussed in Sec.~\ref{sec:WZproduction}. Different contributions to single muon $p_T$ spectra are studied in Sec.~\ref{sec:SingleMuon}. The ALICE capabilities for W/Z detection via their muonic decay in the muon spectrometer are presented in Sec.~\ref{sec:Wdetection} and \ref{sec:Zdetection} respectively. Muon charge asymmetry and a possible application for the study of medium induced effects in Heavy Ion Collisions (HIC) are examined in Sec.~\ref{sec:Other}.

\section{Production of W and Z bosons at LHC}
\label{sec:WZproduction}

W/Z bosons have been simulated by means of the PYTHIA 6.2 Monte-Carlo event generator \cite{pythiadoc} with CTEQ4L PDFs \cite{cteq4} in the AliRoot framework \cite{AliRoot}. 
Yields in A-A and p-A collisions have been obtained from the PYTHIA simulations of nucleon-nucleon collisions, assuming binary scaling and the EKS98 shadowing parameterization \cite{eks98}. 
 Default PYTHIA parameters for studies of W/Z inclusive generation have been used. That is, considering $2 \rightarrow 1$ processes and parton showers (initial and final state radiation). In references \cite{pythiadoc,miu} it is shown that in this way the Tevatron W $p_T$ spectrum is reproduced.
\begin{figure*}[!htb]  
  \centering
  \begin{minipage}[c]{.49\linewidth}
    \centering\includegraphics[width=\columnwidth]{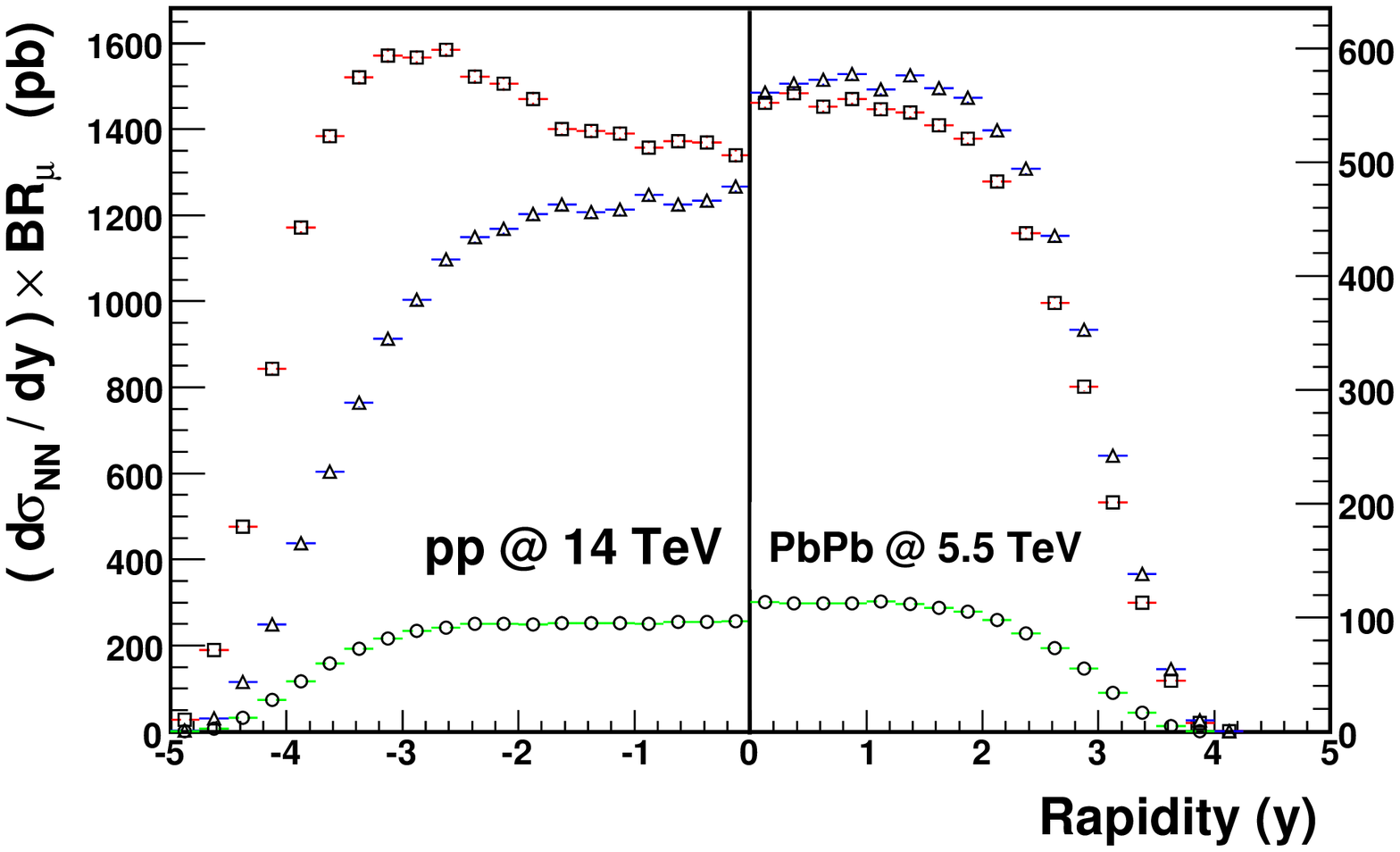}
  \end{minipage}
  \begin{minipage}[c]{.49\linewidth}
    \centering\includegraphics[width=\columnwidth]{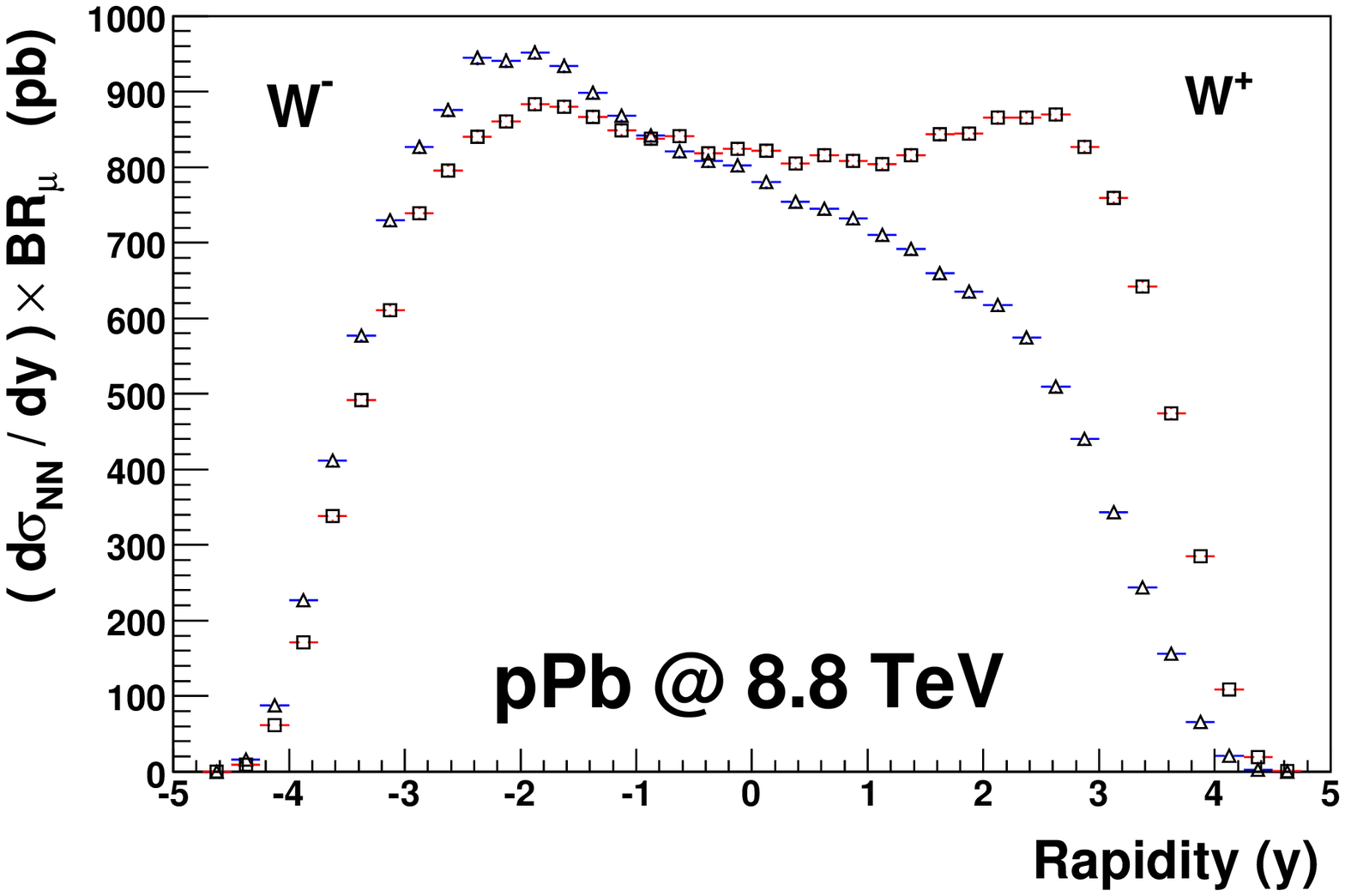}
  \end{minipage}
  \caption{\label{fig:WGenerated_y}W/Z nucleon-nucleon differential production cross-section (multiplied by the muonic branching ratio) as a function of rapidity. On the left plot, the left half correspond to the distributions for p-p collisions at $14$~TeV, and the right half to Pb-Pb collisions at $5.5$~TeV. On the right plot the distributions correspond to the p-Pb case at $8.8$~TeV. Open squares represent W$^+$, open triangles W$^-$, and open circles Z$^0$.}      
\end{figure*}
\newline
The W/Z differential production cross-section (times the muonic branching ratio) is presented in Fig.~\ref{fig:WGenerated_y} as a function of rapidity. The left figure corresponds to p-p collisions at $14$~TeV (left half) and Pb-Pb collisions at $5.5$~TeV (right half), and the right figure to p-Pb collisions at $8.8$~TeV. 
The isospin asymmetries of the colliding systems induce charge asymmetries on the W production cross-section. These charge asymmetries are observed in Fig.~\ref{fig:WGenerated_y} for different collision types. 
\begin{figure}
  \centering
  \resizebox{0.98\columnwidth}{!}{\includegraphics{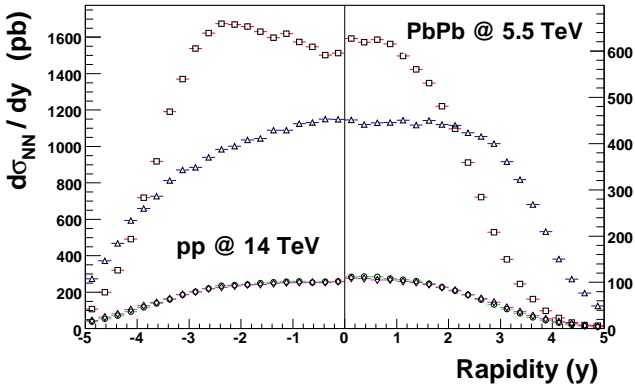}}
  \caption{\label{fig:MuonGen_y_ppPbPb}Muon differential production cross-section per nucleon-nucleon collision from W/Z decays as a function of rapidity for p-p collisions at $14$~TeV (left half) and Pb-Pb collisions at $5.5$~TeV (right half). Open squares (open triangles) represent W $\mu^+$ ($\mu^-$) decays, and open circles (open diamonds) $\mu^+$ ($\mu^-$) from Z.}
\end{figure}
\newline
The vector-axial nature of the $\mu$-W coupling causes an angular asymmetry in W decays.
The predicted muon differential production cross-section from W decays is presented in Fig.~\ref{fig:MuonGen_y_ppPbPb} as a function of rapidity.

\section{Single muon $p_T$ spectra}
\label{sec:SingleMuon}

\begin{figure*}[!htb]  
  \centering
  \resizebox{0.7\linewidth}{!}{\includegraphics{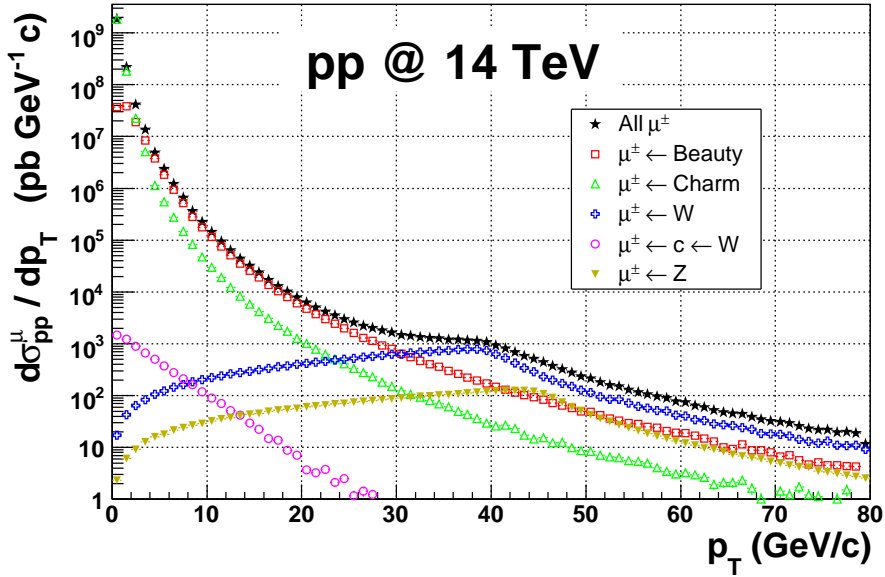}}
  \caption{\label{fig:MuonGen_Pt_pp}Differential muon production cross-section as a function of transverse momentum for p-p collisions at $14$~TeV in $4~\pi$. }
\end{figure*}

The feasibility of W/Z detection via single muons is analysed in this section. 
The $p_T$ distribution of muons from W/Z decays is peaked around one half the W/Z mass. Therefore, we shall concentrate on the high $p_T$ part of the spectrum.
\newline
As single muon sources at LHC we consider:
\begin{itemize}
\item[-] Decays of light flavour and strange hadrons: pions, kaons, \dots 
 These populate the low $p_T$ region \cite{alicepprII}, $p_T \leq 2-3$~GeV/c, and they are neglected in this study. 
\item[-] Open and hidden charm decays (e.g. $D \rightarrow \, l \, {\rm anything } \; ,$ $J/\psi \rightarrow \, \mu^+ \, \mu^- $).
\item[-] Open and hidden beauty decays (e.g. $B  \rightarrow \,  \, l \, \nu_l \, {\rm anything } \; ,$ $\Upsilon \rightarrow \, \mu^+ \, \mu^- $). 
\item[-] W/Z decays: $ W^{+} \rightarrow \; \mu^{+} \; \nu_{\mu} \; ,$ $Z^{0} \rightarrow \; \mu^{+} \;  \mu^{-} \; ,$ $W \rightarrow \; c \, X \rightarrow \; \mu \, {\rm anything} $.
\end{itemize}
Heavy quark production has been simulated by means of PYTHIA. Distributions have been tuned in order to reproduce NLO pQCD results, of order $O(\alpha_s^3)$ \cite{alicepprII}. W/Z production has been simulated as explained in Sec.~\ref{sec:WZproduction}. Nuclear modification of the PDFs have been taken into account in p-A and A-A collisions according to EKS98 parameterization \cite{eks98}. 
\begin{table}
\caption{Charm, beauty, W and Z production cross-sections per nucleon-nucleon collision from NLO calculations. Shadowing is included in Pb-Pb and p-Pb calculations.}
\label{tab:cross_section}   
\begin{tabular}{cccc}
\hline\noalign{\smallskip}
 collision ($\sqrt{s_{NN}} \, [TeV]$)& p-p (14) & p-Pb (8.8) & Pb-Pb (5.5) \\\noalign{\smallskip}\hline\noalign{\smallskip}
 $\sigma_{NN}^{c\bar{c}}$ [mb] & 11.2~\cite{alicepprII} & 7.16~\cite{alicepprII} & 4.32~\cite{alicepprII} \\\noalign{\smallskip}
 $\sigma_{NN}^{b\bar{b}}$ [mb] & 0.51~\cite{alicepprII} & 0.27~\cite{alicepprII} & 0.18~\cite{alicepprII} \\\noalign{\smallskip}
 $\sigma_{NN}^{W} \, \times BR_{\mu \nu}$ [nb]        & 20.9~\cite{frixione_mangano} & 11.3~\cite{vogt} & 6.56~\cite{vogt} \\\noalign{\smallskip}
 $\sigma_{NN}^{Z} \, \times BR_{\mu^+ \mu^-}$ [nb]        & 1.9~\cite{tricoli} &  1.1~\cite{vogt}  &  0.63~\cite{vogt} \\
\noalign{\smallskip}\hline
\end{tabular}
\end{table}
\newline
The expected values of charm, beauty, W and Z production cross-sections per nucleon-nucleon collision are summarized in table \ref{tab:cross_section}. 
The expected single muon differential production cross-section in $4~\pi$ is presented in Fig.~\ref{fig:MuonGen_Pt_pp} as a function of transverse momentum for p-p collisions. 
We note that in addition to W leptonic decays, muons can also be produced via W charmed decays ($W \, \rightarrow \, c \, X \, \rightarrow \,\mu \,\textrm{anything}$). 
However, the latter contribution (represented by open circles in the figure) populates the low $p_T$ region, and hence is of little use for W detection. 
Open charm and beauty decays spread over a wide $p_T$ domain. For $p_T$ larger than $4$~GeV/c, the beauty contribution prevails over the charm contribution. For $p_T$ larger than $30$~GeV/c, muons from W leptonic decays are the main contributors to the spectrum. 
The Z bosons production cross-section (multiplied by the muonic branching ratio) is ten times lower than the W one. 
\newline
Similar calculations for Pb-Pb collisions show the same behaviour. In conclusion, we expect the W boson signal to be visible in the single muon $p_T$ distribution.

\section{ALICE capabilities for W detection in the muon spectrometer}
\label{sec:Wdetection}

The muon spectrometer covers a pseudo-rapidity range $-4.0 < \eta < -2.5$ ($171^{\circ} < \theta < 178^{\circ}$) and has full azimuthal coverage \cite{alicepprII,alicepprI}. 
It is composed of a front absorber, a beam shield, the tracking chambers, a dipole magnet, a muon filter and the trigger chambers. The front absorber and the beam shield reduce the tracking chambers background due to photons and hadrons from the interaction vertex. The tracking system is formed of 10 planes of Cathode Pad Chambers (CPC) with high granularity. Four planes are placed before, two inside and four after the dipole magnet. Tracking inside the $0.7$~T magnetic field of the dipole allows the determination of the muon momenta. The trigger detector is equiped with an iron muon filter, to further diminish the background on the trigger chambers, which are made of Resistive Plate Chambers \cite{guerin,guerinAN}. The trigger allows to select events with candidate muon or dimuon tracks above a given $p_T$. 
In our studies the effect of the trigger has been taken into account by applying a software $p_T$ cut at $1$~GeV/c. In addition, a $p$ cut at $4$~GeV/c has been introduced in order to mimic the effect of the muon absorber. 

\subsection{Reconstruction efficiency in the muon spectrometer}

The reconstruction efficiency for single muon tracks has been evaluated as follows. 
A flat muon $p_T$ distribution has been used to evaluate the efficiency in p-p and in peripheral Pb-Pb collisions. Results indicate that the spectrometer is able to reconstruct muons up to $1$ TeV momenta, and that the reconstruction efficiency is approximately flat in $p_T$ and close to $97~\%$ in the $5~<~p_T~<~60$~GeV/c range.
For the case of central Pb-Pb collisions, the efficiency has been estimated by means of a flat muon $p_T$ distribution merged with HIJING central events (in order to account for detector occupancy in central collisions). The corresponding efficiency is about $95~\%$.

\subsection{Expected muon yields from W decays}

\begin{figure*}[!htb]  
  \centering
  \resizebox{0.7\linewidth}{!}{\includegraphics{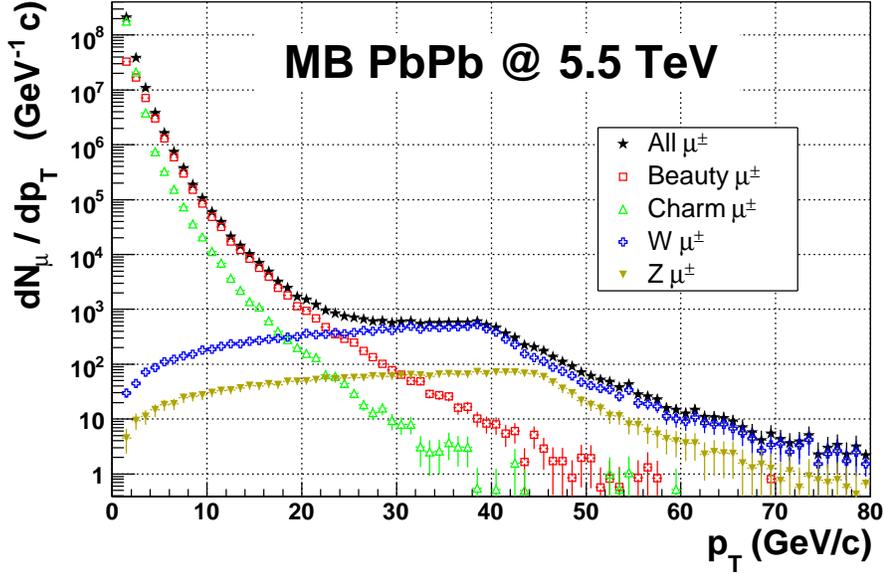}}
  \caption{\label{fig:NumMuonReco_Pt_PbPb}Expected differential number of muons reconstructed in the ALICE muon spectrometer in minimum bias (MB) Pb-Pb collisions in ALICE during one year of data-taking.}
\end{figure*}

Under standard data-taking conditions, during one year of data-taking ALICE expects to accumulate an integrated luminosity of $\cal{L}~=$~$30$~pb$^{-1}$ in p-p collisions, $0.5$~nb$^{-1}$ in Pb-Pb collisions, and $0.1$~pb$^{-1}$ in p-Pb collisions \cite{alicepprI,gines}. 
The muon spectrometer acceptance for muons from W decays has been evaluated to be about $14~\%$, $10~\%$, $17~\%$, $7~\%$ for p-p, Pb-Pb, p-Pb and Pb-p collisions at $14$, $5.5$, $8.8$ and $8.8$~TeV respectively. Table \ref{tab:Expected_Wdecays} presents the estimated number of muons from W decays produced in $4~\pi$ and reconstructed in the muon spectrometer for these conditions in minimum bias (MB) collisions.
\begin{table}
\caption{\label{tab:Expected_Wdecays}Estimated number of muons from W decays in ALICE during one year of data-taking. Statistics in $4~\pi$ ($N_{\mu \leftarrow W}$) and evaluated number of reconstructed muons ($N_{\mu \leftarrow W}^{Reco}$).}
\begin{center}
\begin{tabular}{ccc} \hline  \noalign{\smallskip}  
 Collision  & $N_{\mu \leftarrow W}$  &  $N_{\mu \leftarrow W}^{Reco}$ \\ \noalign{\smallskip}\hline\noalign{\smallskip} 
 p-p at $14$ TeV    &  $6.3 \cdot 10^{5}$  &  $8.6 \cdot 10^{4}$ \\\noalign{\smallskip}
 p-Pb at $8.8$ TeV  &  $2.3 \cdot 10^{5}$  &  $4.0 \cdot 10^{4}$ \\\noalign{\smallskip}
 Pb-p at $8.8$ TeV  &  $2.3 \cdot 10^{5}$  &  $1.7 \cdot 10^{4}$  \\ \noalign{\smallskip}
Pb-Pb at $5.5$ TeV  &  $1.4 \cdot 10^{5}$  &  $1.4 \cdot 10^{4}$ \\ \noalign{\smallskip}\hline\noalign{\smallskip} 
\end{tabular}
\end{center}
\end{table}
\newline
The expected muon yields in p-p and Pb-Pb collisions for two $p_T$ ranges and different centrality classes are presented in table \ref{tab:Expected_Wdecays_ppPbPb}. The extrapolations from nucleon-nucleon to Pb-Pb/p-Pb total cross-sections and the evaluations for different centrality classes are carried out by binary scaling \cite{glauber,denterria}, taking into account the shadowing effects as discussed above. 
The $p_T$ distribution of reconstructed muons from different sources in minimum bias (MB) Pb-Pb collisions is shown in Fig.~\ref{fig:NumMuonReco_Pt_PbPb} for standard data-taking conditions. 
\begin{table}
\caption{\label{tab:Expected_Wdecays_ppPbPb}Estimated number of reconstructed muons from W decays in ALICE during one year of data-taking ($N_{\mu \leftarrow W}^{Reco}$) as a function of the $p_T$ range (in GeV/c) and centrality class (C.C.).}
\begin{center}
\begin{tabular}{cccc} \hline \noalign{\smallskip}  
\multirow{2}*{Collision} & \multirow{2}*{C.C.} &  $N_{\mu \leftarrow W}^{Reco}$  &  $N_{\mu \leftarrow W}^{Reco}$ \\ 
  &  &  $p_T \in (0,80)$ &  $p_T \in (30,50)$ \\ \noalign{\smallskip}\hline\noalign{\smallskip} 
p-p at 14 TeV & MB  &  $8.6 \cdot 10^{4}$  &  $5.0 \cdot 10^{4}$ \\ \noalign{\smallskip}\hline\noalign{\smallskip}  
  & MB &  $1.4 \cdot 10^{4}$  &  $6.7 \cdot 10^{3}$  \\ 
\multirow{2}*{ Pb-Pb } & 0-5 \% &  $3.4 \cdot 10^{3}$  & $1.6 \cdot 10^{3}$  \\ 
  & 0-10 \% & $6.0 \cdot 10^{3}$  & $2.9 \cdot 10^{3}$  \\ 
\multirow{1}*{ at 5.5 TeV } 
  & 40-70 \% & $1.0 \cdot 10^{3}$ & $4.9 \cdot 10^{2}$  \\ 
  & 50-70 \% & $4.2 \cdot 10^{2}$ & $2.0 \cdot 10^{2}$   \\ \noalign{\smallskip}\hline\noalign{\smallskip} 
\end{tabular}
\end{center}
\end{table}

\section{Perspectives for Z detection in the muon spectrometer}
\label{sec:Zdetection}

Z bosons could be reconstructed by invariant mass analysis of dimuon pairs. Preliminary studies in the AliRoot framework \cite{AliRoot} show that the spectrometer acceptance is about $4.5~\%$~($1.7~\%$) in p-p collisions at $14$~TeV (in Pb-Pb collisions at $5.5$~TeV) with respect to $4~\pi$ \cite{blusseau}. It has been estimated that during one year of data-taking about $2500$~($230$) dimuons from Z decays will be produced in the spectrometer acceptance for p-p~(Pb-Pb) MB collisions.
These studies indicate that Z measurements in p-p collisions in the muon spectrometer are feasible, but in the Pb-Pb case statistics from several runs should be accumulated. 
Invariant mass resolution has been evaluated to be approximately $2~\%$ in p-p collisions.

\section{Physics applications}
\label{sec:Other}

The characteristics of electroweak processes can be exploited to obtain a measurement of W production, and can be used as a reference for observing medium induced effects on other probes.

\subsection{Muon charge asymmetry}

\begin{figure}
  \centering
  \resizebox{0.98\columnwidth}{!}{\includegraphics{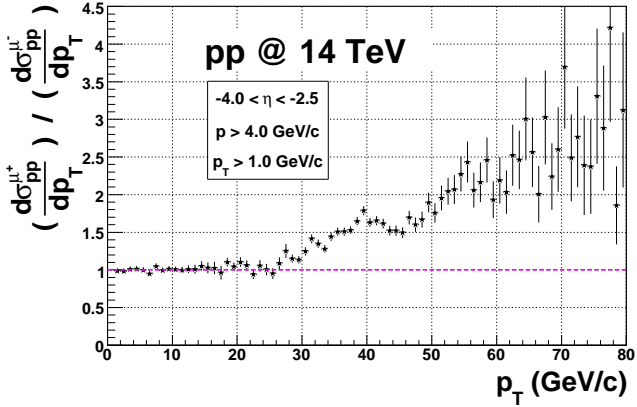}}
  \caption{\label{fig:MuonFracAcc_Pt_pp}$\mu^+ \, / \mu^-$ ratio in the muon spectrometer acceptance for p-p collisions at 14 TeV.}
\end{figure}

\begin{figure}
  \centering
  \resizebox{0.98\columnwidth}{!}{\includegraphics{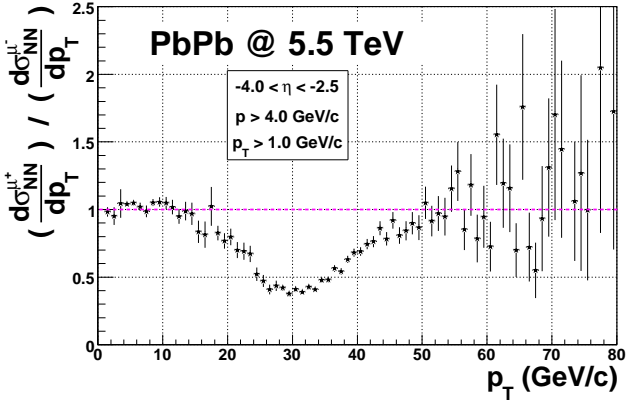}}
  \caption{\label{fig:MuonFracAcc_Pt_PbPb}$\mu^+ \, / \mu^-$ ratio in the muon spectrometer acceptance for Pb-Pb collisions at 5.5 TeV.}
\end{figure}

One of the peculiarities of W production is the induced asymmetry on positive and negative muon yields (muon charge asymmetry).
Figures~\ref{fig:MuonFracAcc_Pt_pp} and \ref{fig:MuonFracAcc_Pt_PbPb} present the $\mu^+ \, / \mu^-$ production cross-section ratio as a function of $p_T$ for p-p and Pb-Pb collisions respectively. Charm, beauty, W and Z contributions have been taken into account. 
As expected, the $\mu^+ \, / \mu^-$ ratios differ from unity for $p_T$ larger than $30$~GeV/c (where the contribution from W leptonic decays is dominant), and their shape depends on the collision type.

\subsection{Sensitivity to heavy quark high $p_T$ suppression}

In relativistic HIC, a dense and hot state of matter is expected to be formed \cite{karsch}. 
Heavy quark and quarkonia production are sensitive probes of the formation of this state of matter in HIC \cite{satz}. In particular, a suppression of quarkonia production in HIC due to color screening and an energy loss of heavy quarks caused by gluon radiation are expected \cite{satz,andrea,andrea_proc}. 
Since electroweak bosons do not interact strongly, they could be used as a reference. For instance, single muon $p_T$ distribution is mainly populated by heavy quark production for $10~<~p_T~<~25 $ GeV/c and by electroweak bosons for $p_T~>~30$ GeV/c. We define a parameter, ${\cal S}$, as the ratio of single muon yields in two $p_T$ ranges, e.g. the ratio between $(15,20)$~GeV/c and $(30,40)$~GeV/c:
\begin{equation}
{\cal S} =  \frac{N^{\mu}_{15-20}}{N^{\mu}_{30-40}}
\end{equation}
${\cal S}$ will then carry information on nuclear effects (both initial and final). It will then be important to evaluate the contribution of initial state effects such as shadowing by studying pA collisions. 
\newline
We have estimated that ${\cal S} \approx \, 4.7$ in absence of the final state effects for MB Pb-Pb collisions. If energy loss scenarios are verified, a reduction factor in the order of $2-4$ could be expected in Pb-Pb central collisions \cite{alicepprII} with respect to theoretical calculations.

\section{Summary}
\label{sec:Summary}

The production of electroweak bosons can be measured with the ALICE muon spectrometer. They should provide a useful tool for luminosity measurements and to improve the evaluation of the detector performances. In the muon spectrometer, the single muon charge asymmetry can be used to obtain a measurement of W production. W detection in the spectrometer in p-A collisions will provide unique information on the quark nuclear modification effects. Unique because ALICE will be the only experiment at LHC covering the high rapidity region in p-A/A-A collisions. In A-A collisions, electroweak bosons detection will permit to check the binary scaling and could be used as a reference for heavy quark high $p_T$ suppression.

\begin{acknowledgement}
The author would like to thank G. Martinez Garcia, who actively participated in this work.
The author would also like to thank F.~Antinori, L.~Aphecetche, A.~Bellingeri, Ph.~Crochet, Ch.~Finck, T.~Gousset, E.~Vercellin and ALICE Heavy Flavours Working Group (PWG3) for fruitful discussions.

Part of this work was supported by the EU Integrated Infrastructure
Initiative HadronPhysics Project under contract number
RII3-CT-2004-506078.
\end{acknowledgement}



\end{document}